\documentclass[a4paper,11pt]{article}
\pdfoutput=1
\usepackage{jheppub}
\usepackage{amsmath,amssymb,tabularx,graphicx,cancel,subcaption,xspace,flafter,ifpdf,todonotes,slashed,cleveref,hyperref,booktabs}
\usepackage{placeins}
\usepackage[force]{feynmp-auto}
\def\spa#1.#2{\left\langle#1\,#2\right\rangle}
\def\spb#1.#2{\left[#1\,#2\right]}
\def\spaa#1.#2.#3{\langle\mskip-1mu{#1}
                  | #2 | {#3}\mskip-1mu\rangle}
\def\spbb#1.#2.#3{[\mskip-1mu{#1}
                  | #2 | {#3}\mskip-1mu]}
\def\spab#1.#2.#3{\langle\mskip-1mu{#1}
                  | #2 | {#3}\mskip-1mu\rangle}
\def\spba#1.#2.#3{\langle\mskip-1mu{#1}^+
                  | #2 | {#3}^+\mskip-1mu\rangle}
\def\spav#1.#2.#3{\|\mskip-1mu{#1}
                  | #2 | {#3}\mskip-1mu\|^2}
\def\jc#1.#2.#3{j^{#1}_{#2#3}}
\def\blfootnote{\xdef\@thefnmark{}\@footnotetext}

\newcommand{\madgraph}{\texttt{MadGraph5\_aMC@NLO}\xspace}
\newcommand{\dd}{\ensuremath{\mathrm{d}}}
\preprint{\begin{minipage}[t]{\widthof{MCNET-20-14,}}
    DCPT/20/30\\
    DESY 20-090\\
    IPPP/20/15\\
    LU-TP-20-21\\
    MCNET-20-14\\
    SAGEX-20-12
  \end{minipage}}
\title{A Positive Resampler for Monte Carlo Events with Negative Weights}
\author[a]{Jeppe~R.~Andersen,}
\author[b]{Christian G\"utschow,}
\author[c]{Andreas Maier}
\author[d]{and Stefan Prestel}

\emailAdd{jeppe.andersen@durham.ac.uk}
\emailAdd{chris.g@cern.ch}
\emailAdd{andreas.martin.maier@desy.de}
\emailAdd{stefan.prestel@thep.lu.se}

\affiliation[a]{Institute for Particle Physics Phenomenology, University of Durham, Durham, DH1 3LE, UK}
\affiliation[b]{Department of Physics and Astronomy, University College
  London,\\Gower Street, London, WC1E 6BT, UK}
\affiliation[c]{Deutsches Elektronen-Synchrotron, DESY, Platanenallee 6,
  15738 Zeuthen, Germany}
\affiliation[d]{Theoretical Particle Physics, Department of Astronomy and Theoretical Physics, Lund University,
S{\"o}lvegatan 14 A, SE-223 62 Lund, Sweden}

\abstract{We propose the \emph{Positive Resampler} to solve the problem
  associated with event samples from state-of-the-art predictions for
  scattering processes at hadron colliders typically involving a sizeable
  number of events contributing with negative weight. The proposed method
  guarantees positive weights for all physical distributions, and a
  correct description of all observables. A desirable side product of the
  method is the possibility to reduce the size of event samples produced
  by General Purpose Event Generators, thus lowering the resource demands for
  subsequent computing-intensive event processing steps.
  We demonstrate the viability and
  efficiency of our approach by considering its application to a
  next-to-leading order + parton shower merged prediction for the production of
  a $W$ boson in association with multiple jets.}

\begin{document}
\maketitle
\flushbottom

\section{Introduction}
\label{sec:intro}
General Purpose Event
Generators~\cite{Bellm:2015jjp,Sjostrand:2014zea,Bothmann:2019yzt} form a
crucial component of studies in high-energy physics, since they produce
detailed predictions used for the design and calibration of detectors,
interpretations of the measurements as well as the investigations of theoretical models.
More often than not it is necessary to take into account the effects from the
perturbative showering and the hadronisation models implemented in these
generators, in order to achieve an accurate prediction for the cuts and
observables chosen for experimental measurements.

High-accuracy perturbative event generator predictions
can be obtained by first matching each jet multiplicity to next-to-leading order (NLO)
using the methods of e.g.~MC@NLO~\cite{Frixione:2002ik} or
POWHEG~\cite{Nason:2006hfa}, followed by a merging of these exclusive samples
using approaches such as MEPS@NLO~\cite{Hoeche:2012yf} or
UNLOPS~\cite{Lonnblad:2012ix,Lonnblad:2012ng}. The increased accuracy comes at a
significant cost in additional computing resources, and these calculations
increasingly contribute to the LHC computing footprint. The result of these
merged NLO-accurate event generator simulations are event samples
containing events of both positive and negative
weights, meaning that the correspondence between the number of events in a
bin of a distribution and the cross section in that bin is
lost.\footnote{The fraction of negative weights can vary wildly between
different matching or merging schemes, but in general will be non-negligible
for processes containing multiple light jets. This is illustrated on
Figures~\ref{fig:distributions}-\ref{fig:pt123}.} Even when
the event samples are unweighted to constitute events with weights of $\pm$1,
the number of negative weighted events can be significant. This reduces the
statistical significance of the sample compared to one with the same number
of all positive weight events.

The outcome of event generation is often processed through time-consuming
detector simulation -- which currently constitutes the major part of the LHC
computing budget. Since both positive and negative weight events are afterwards
processed at a significant cost, it is beneficial to reduce the cancellation
of events of negative and positive weight. This can be done by reducing the
occurrence of negative weight events in the event generation, see
e.g.~\cite{Frederix:2020trv} for an approach in MC@NLO.

We report here on an alternative and NLO-matching-independent approach to
completely remove all negative weights from any already generated
sample, and re-introducing the correspondence between the number of events in
a bin and the local contribution to the cross section. This \emph{positive
  resampling} will be achieved in a two-stage process: 1) modify the weights
of events to be all positive and possibly smaller in magnitude, and 2) apply
a standard unweighting. The second step should be taken only if the number of
events is sought to be reduced. Reducing the event sample can
significantly lower the computing budget for the steps in the analysis
following the event generation, both in terms of CPU and disk. It may seem
counter intuitive to allow for or even seek a reduction in the number of
events, since traditionally the statistical significance of a sample, or the
variance, is linked to the number of events, and reducing the number of
events would therefore reduce the statistical significance of the
sample. However, when the sample contains events with both positive and
negative weights, the number of events can indeed be reduced without
impacting the statistical significance: The effective cross section is given by
$\sigma=\sigma_p-\sigma_n$, where $\sigma_p$ is the contribution from events
with positive weights, and $\sigma_n$ that from negative weights. The
Monte Carlo variance associated with the sample is
$s_p^2/N_p+s_n^2/N_n$, where $s_p^2, s_n^2$ is the variance of the
integrand. If we replace the sample with one that has
$N_s$ positively weighted events, the variance of this new sample is
$s_s^2/N_s$. If therefore $N_s/(N_p+N_n)$ is similar to
$s_s/s_p$ or $s_s/s_n$, the variance can be unchanged. This is achieved with the
\emph{Positive Resampler}, a simplified description of which transforms the
weight of each event to its absolute value and multiplies by
$(\sigma_p-\sigma_n)/(\sigma_p+\sigma_n)$. If there is a large cancellation,
then the weight of each event is much reduced, leading also to a reduction in
the Monte Carlo estimate of the variance. Hence fewer events are needed for
the same statistical certainty. The
\emph{Positive Resampler} is introduced in section~\ref{sec:posres},
and section~\ref{sec:results} showcases results obtained based on samples of
$W\!+\!0,1,2$ jets at NLO fixed-order accuracy merged with UNLOPS.

\section{Weights in UNLOPS merging}
\label{sec:unlops}

High-precision predictions are important ingredients to LHC data
analysis. If the analysis is sensitive to the effect of yet higher jet
multiplicities, high precision is obtained by ``merging" several distinct
calculations. NLO merged calculations provide the state-of-the-art for LHC
phenomenology, and contribute significantly to the overall computing resource
usage. At the same time, increased precision almost always comes at the cost
of having to rely more heavily on weighted event generation. Typically,
issues due to a reduction in statistical convergence worsen
with every additional NLO calculation included in the merging.

NLO merging schemes aim to produce
inclusive event samples that both comply with  NLO fixed-order accuracy for
several multi-jet processes, and ensure that the accuracy of parton showering
is preserved. Various methods have been developed to this
effect~\cite{Hoeche:2012yf,Lonnblad:2012ix,Platzer:2012bs,Hamilton:2012rf,
Alioli:2012fc,Frederix:2012ps,Bellm:2017ktr}, all differing slightly
in the concrete goals and the definition of the target accuracy. Each NLO
merging method suffers from various sources of negative weights.
Predictions based on unitarised merging schemes~\cite{Lonnblad:2012ix} are
the only predictions that not only comply with both of the above criteria,
but also guarantee that, for an arbitrary base process and
\emph{arbitrary} parton multiplicity, inclusive $n$-jet cross sections are preserved
exactly, without introducing sub-dominant contributions due to the merging.\footnote{
It should however be noted that calculations that are
valid for specific processes~\cite{Alioli:2012fc,Alioli:2019qzz}, or up to a maximal
multiplicity~\cite{Hamilton:2012rf,Frederix:2015fyz} also fulfill similar consistency criteria.}
This desirable feature comes at the price of introducing new sources
of counter-events and/or event weights compared to other methods such
as~\cite{Hoeche:2012yf}.

Hence, unitarised NLO merging provides a very non-trivial test of unweighting
methods. In the following, we will employ the NLO merging
as implemented in the \textsc{Dire} plugin to the \textsc{Pythia} event
generator as a test case. This NLO merging implementation is based on
UNLOPS~\cite{Lonnblad:2012ix}, includes QCD, QED and electroweak vector boson emissions, and thus
allows merging of calculations with multiple hard jets,
photons/leptons or electroweak vector bosons. In particular, the ``EW-improved"
merging of~\cite{Christiansen:2015jpa} is extended to NLO QCD. Unordered configurations
(i.e.\ for which no history of ordered emissions can be reconstructed
for an input event) are treated according to the MOPS+unordered
prescription of~\cite{Fischer:2017yja}. The latter means that the merging
also heavily relies on matrix elements extracted from \madgraph~\cite{Alwall:2014hca}.
Within this NLO merging framework, several sources of event weights arise:
\begin{itemize}
\item The input NLO QCD short-distance cross sections (generated with \madgraph for this
study) can contain positive and negative weights. The input events are
typically unweighted to $\pm w$, where $w$ is a unit weight.
\item The procedure of assigning a parton shower history to
a high-multiplicity input event can necessitate weights. The history is
chosen among all possible shower histories according to the
shower probability, which may contain non-positive definite splitting
functions in \textsc{Dire}. This leads to a corrective (positive or negative)
weight.
\item The merging procedure enforces a consistent renormalisation and
factorisation scale setting by introducing weights. These weights are
almost entirely positive and tend to fluctuate only mildly. The merging scheme
further removes overlap between different input samples by including
no-emission probabilities. These factors are essentially in $[0,1]$, but can,
in rare cases (e.g.\ due to negative NLO parton distribution functions or splitting functions), lead
to negative weights.
\item The $\mathcal{O}(\alpha_s)$ expansion of the weight discussed in the
previous point -- which is necessary to guarantee the NLO accuracy of the
method -- is often negative, thus introducing a non-negligible source
of weights.
\item The accuracy of inclusive cross sections is enforced by explicit
unitarisation~\cite{Lonnblad:2012ng,Lonnblad:2012ix}. This means that a large fraction of
events will be employed as ``counter-events" with negative weight.
\item Independently of NLO merging, the subsequent showering may also
generate event positive or negative weights if the splitting functions
are not positive-definite.
\end{itemize}

The regions of phase space where negatively weighted events contribute most
depends on the source of weights, and thus also on the merging setup. An
overall larger fraction of negative weights can be expected when merging
more NLO calculations. This is best illustrated with an example. Consider
a precise background prediction for a vector-boson $+$ jets measurement.
If only the inclusive zero-jet prediction is NLO accurate, then the
predominant source of negative weights is the unitarisation of reweighted
one-jet LO configurations. This will lead to a moderate amount of
negatively weighted events, which produce a relatively flat negative
contribution to the vector boson rapidity spectrum. The same events will induce
larger negative weight fraction at small boson $p_{\perp V}$ than at
high $p_{\perp V}$, since showering from counter-events is constrained to the
soft/collinear regions. Negative weights will have a negligible impact
on observables that require one, two or more jets.
If the calculation is extended by also
including an NLO calculation for the inclusive $V+$one-jet rate, then
$p_{\perp V}$ will also exhibit negative-weight contributions at high values,
since new sources of negative weights ($\mathcal{O}(\alpha_s)$ expansions,
unitarisation of the two-jet LO sample) arise. Observables that depend
on two jets will now acquire a negative component at small jet separation due
to the mechanism of unitarisation. This effect is illustrated in Figure~\ref{fig:distributions}.

\section{A Positive Resampler}
\label{sec:posres}

Our main goal is to restore the connection seen when generating all positive
weight events between the number of events in the neighbourhood of any phase
space point (e.g.~the bin in any histogram) and the local contribution to the
cross section. This will require a modification of the weight of all events in
the neighbourhood of events with negative weights. The idea is simply stated
to 1) calculate the cross section $\sigma_i$ from the $N_i$ positive
and negative weight events in a neighbourhood, 2) change each weight
to be positive and rescale all weights to preserve $\sigma_i$. One can
then proceed with a unweighting procedure over all
the neighbourhoods $i$ to restore the connection between the number of
events and the cross section from each neighbourhood.

In Monte Carlo event generation, the cross sections are generated
exclusively in all momenta. We will demonstrate first the method for the
idealised situation where the cross section is stored
differentially in
all momenta relevant for the later event analysis -- this could be
e.g.~momenta of the jets, leptons etc. We will then demonstrate how it works
also in the case of using simple binned observables for the unweighting.

\subsection{Multi-Dimensional Resampling}
\label{sec:multidimsampl}
We begin by considering the idealistic case (i.e.~the limit of infinitely
narrow bin widths) where a differential distribution in the observable
$\mathcal{O}_1$ is calculated with both positive and negative weight events
in a $f$-body phase space region $\Omega_f$. The differential distribution in
the observable $\mathcal{O}_1$ is then constructed as
\begin{align}
\label{eq:diff1}
  \frac{\dd \sigma}{\dd \mathcal{O}_1} = \frac{\dd
  \sigma}{\dd\Omega_f}\ \frac{\dd \Omega_f}{\dd \Omega_n}\ \frac{\dd\Omega_n}{\dd\mathcal{O}_1},
\end{align}
where
\begin{itemize}
\item[$\frac{\dd\sigma}{\dd\Omega_f}$]  signifies the cross section calculated
  in terms of the final state momenta.
\item[$\frac{\dd \Omega_f}{\dd \Omega_n}$] encodes e.g.~the jet clustering
  and is the Jacobian for the $f$-body phase space into the $n$-body phase
  space ($n<f$) that the observable depends on.
\item[$\frac{\dd\Omega_n}{\dd\mathcal{O}_1}$] is traditionally included in
  the calculations by the binning of the $n$-body phase space in terms of
  $\mathcal{O}_1$.
\end{itemize}
%% Unweighting clearly describes all observables O_1 if event sample can be constructed in omega_f or omega_n
Equation~\eqref{eq:diff1} just constitutes the chain rule of
differentiation. Histogrammed distributions are obtained by simple
integrations of this relation. It is therefore not surprising that if the
events of positive and negative weights arising in the generation
(represented by $\frac{\dd\sigma}{\dd\Omega_f}$) can be turned into events of
all positive weights, then the distribution in any $\mathcal{O}_1$ can be
calculated with these events (since the calculation of the other factors in
equation~\eqref{eq:diff1} are unchanged). Of course the $f$-body phase space
can have a very high dimension and it may seem impractical to perform the
clustering of events in bins in all the dimensions of $\Omega_f$ followed by
the reweighting procedure outlined above. One could of course
perform the reweighting in the lower-dimensional $\Omega_n$, the phase space of all objects
(jets etc.) entering observables. It is clear here that the reweighting
works for any observable $\Omega_1$. What is perhaps surprising is that if
the reweighting is performed in neighbourhoods (bins after integration) in
$\mathcal{O}_1$ then $\frac{\dd\sigma}{\dd \mathcal{O}_2}$ for any other
observable $\mathcal{O}_2$ can still be constructed: We have
\begin{align}
  \frac{\dd\sigma}{\dd
  \mathcal{O}_2}=\frac{\dd\Omega_f}{\dd\mathcal{O}_2}\ \left\{\left(
  \frac{\dd \sigma}{\dd\Omega_f}\
  \frac{\dd\Omega_f}{\dd\Omega_n}\ \frac{\dd\Omega_n}{\dd\mathcal{O}_1}\
  \right)\ \frac{\dd\mathcal{O}_1}{\dd\Omega_f}\right\}.
\end{align}
All reference to the cross section $\sigma$ is within the brackets
$(\cdots)$, so the cancellation of positive and negative weight events can be
implemented here in terms of reweighting as above to the distribution in
$\mathcal{O}_1$. The effect of
$\dd\mathcal{O}_1/\dd\Omega_f$ is taken into account by calculating the value
(or bin) of $\mathcal{O}_1$ starting from the phase space points in
$\Omega_f$. The spectrum for the observable $\mathcal{O}_2$ is then
calculated by constructing the quantities
\begin{itemize}
\item[$\frac{\dd\Omega_f}{\dd\mathcal{O}_2}$] by binning the contribution
  from the phase space points in $\mathcal{O}_f$ in terms of the observable
  $\mathcal{O}_2$
\item[$\{\cdots\}$] by finding all $\Omega_f$ resulting in $\mathcal{O}_1$,
  and multiply by the differential distribution in $\mathcal{O}_1$.
\end{itemize}
So, in terms of truly differential distributions, it would not matter which
observable $\mathcal{O}_i$ one would start from. The distribution in
$\mathcal{O}_j$ can be obtained by the procedure above. This is correct up to
effects in the binning in $\mathcal{O}_1$, which we will later show are
modest indeed, and can be reduced further by decreasing the bin widths used
in the Positive Resampler. Indeed, in the situation where the impact of the
negative weight events in phase space $\Omega_f$ mapped into the equivalent
bins in $\mathcal{O}_1$ and $\mathcal{O}_2$ is identical, then the result for
$\mathcal{O}_2$ using the Positive Resampler in $\mathcal{O}_1$ is exact. We
will see in section~\ref{sec:PosResHighDim} how convergence can be achieved
by sampling in multiple dimensions.

\subsection{Resampling the total cross section}
\label{sec:posres_xs}

The extreme opposite of resampling in all $n$ dimensions of the momenta of
constructs used in the analysis, such as jets and leptons, is resampling just
the cross section -- effectively using just one bin. As we will see in
section~\ref{sec:results} even this extreme yields reasonable results using
the method described in this section. It may be considered coarse, but we
include the discussion here since it provides a simple example of the
algorithms used. For clarity, we choose to illustrate
the method in terms of bins and weights, obtained as integrals and MC
sampling of the distributions discussed in section~\ref{sec:multidimsampl}.

Let us start by considering the total cross section
\begin{equation}
  \label{eq:sigma}
  \sigma = \sum_{i=1}^N w_i\,,
\end{equation}
obtained from $N$ events with weights $w_i$. Introducing a convenient factor of one, we can write
\begin{equation}
  \label{eq:pos_resample_sigma}
  \sigma = \sum_{i=1}^N w_i = \frac{\left(\sum_{i=1}^N |w_i|\right)\left(\sum_{i=1}^N w_i\right)}{\sum_{i=1}^N |w_i|} \equiv P \sum_{i=1}^N|w_i|\,,
\end{equation}
where $0 \leq P = \tfrac{\sum_i w_i}{\sum_i |w_i|} \leq
1$. Effectively, this amounts to replacing each event weight $w_i$ by
$P|w_i|$. The total cross section is preserved by construction, but
what is the effect on binned distributions?

To answer this question, let us select an arbitrary distribution
$\tfrac{d\sigma}{d\mathcal{O}}$ and an arbitrary bin $B$ ranging from
$\mathcal{O}_B$ to $\mathcal{O}_{B+1}$ and containing $M \gg 1$
events. Without loss of generality we can assume that the bin contains
the events $i=1,\dots,M$. The height of the bin is given by
\begin{equation}
  \label{eq:bin_height}
  \sigma_{\mathcal{O},B} \equiv \frac{d\sigma}{d\mathcal{O}}(\mathcal{O}_{B+1} - \mathcal{O}_B) = \sum_{i=1}^M w_i = \frac{\sum_{i=1}^N |w_i|}{\sum_{i=1}^N |w_i|} \sum_{i=1}^M w_i\,,
\end{equation}
where we have introduced the same factor of one as
previously. In the simplified case with a uniform distribution of negative weights,\footnote{
  This follows directly from the law of large numbers $\frac{1}{N}\sum_{i=1}^N x_i \approx \frac{1}{M}\sum_{i=1}^M x_i$ for random variables $x_i \in \{w_i, |w_i|\}$, but also holds more generally. For example, generating larger event samples in specific regions of phase space introduces a bias in the event weights but does not affect the following conclusions.
  }
\begin{equation}
  \label{eq:N_to_M_sum}
  \frac{\sum_{i=1}^M w_i}{\sum_{i=1}^M |w_i|} \approx \frac{\sum_{i=1}^N w_i}{\sum_{i=1}^N |w_i|}\,,
\end{equation}
we can effectively swap the summation limits in the numerator of the
equation~(\ref{eq:bin_height}) for the bin height:
\begin{equation}
  \label{eq:pos_resample_bin}
  \sigma_{\mathcal{O},B} = \sum_{i=1}^M w_i = \frac{\sum_{i=1}^N |w_i|}{\sum_{i=1}^N |w_i|} \sum_{i=1}^M w_i \approx \frac{\sum_{i=1}^N w_i}{\sum_{i=1}^N |w_i|} \sum_{i=1}^M |w_i| = P \sum_{i=1}^M |w_i|\,.
\end{equation}
This tells us that simply replacing each event weight $w_i$ by $P |w_i|$
preserves the total cross section \emph{exactly} and also all bin heights in
each distribution in the limit of large statistics, i.e.~within statistical
fluctuations for finite statistics. In practice, negative weights will
not necessarily be distributed uniformly. We will analyse the real-world
performance for the example of W + jets production in
section~\ref{sec:results}.

\subsection{Resampling a distribution}
\label{sec:posres_dist}
For one-dimensional sampling in terms of an observable $\mathcal{O}$, the
method can be adapted to preserve exactly the distribution in
$\mathcal{O}$. The accuracy in other distribution will be determined by the
variation in the impact of negative weight events for the phase space
$\Omega_f$ mapped into each bin in the two distributions. In analogy
to equation~(\ref{eq:pos_resample_sigma}) we introduce separate rescaling
factors $P_{\mathcal{O},B}$ for each bin $B$ containing the events
$i=i_B,\dots,i_{B+1}$:
\begin{equation}
  \label{eq:pos_resample_distr}
  \sigma_{\mathcal{O},B} = \sum_{i=i_B}^{i_{B+1}} w_i = \frac{\sum_{i=i_B}^{i_{B+1}}|w_i|}{\sum_{i=i_B}^{i_{B+1}}|w_i|} \sum_{i=i_B}^{i_{B+1}} w_i  \equiv P_{\mathcal{O},B} \sum_{i=i_B}^{i_{B+1}} |w_i| \,.
\end{equation}
This preserves all bin heights exactly and therefore the full
distribution in $\mathcal{O}$. The total cross section is just the sum
over all bins, which also remains unchanged. In cases where the
systematic variation in the distribution of negative weights is
negligible, eq.~(\ref{eq:N_to_M_sum}) guarantees that all rescaling
factors $P_{\mathcal{O'}, B'}$ for all observables $\mathcal{O'}$ and
bins $B'$ converge to the same value. This implies that all other
distributions remain correct in the limit of large statistics.

It is straightforward to generalise the argument to multi-differential
distributions. For instance, if we resample bins in a
double-differential distribution in $\mathcal{O}$ and $\mathcal{O}'$
then also the respective single-differential distributions in
$\mathcal{O}$ and $\mathcal{O}'$ as well as the total cross section
are preserved exactly. The most extreme case is a differential
distribution in all final-state momenta. In the limit of infinitesimal
bin widths we recover the idealised scenario already described in
section~\ref{sec:multidimsampl}. The
convergence to the correct result can be achieved by appropriately increasing the dimensions used in the
Positive Resampler.

\section{Results from the Positive Resampler}
\label{sec:results}

To demonstrate the practical performance of the algorithm outlined in
section~\ref{sec:posres} we consider an $\alpha_s$-driven NLO-merged description of the
$W$-boson $+$ jets process, as an example of a resource-intense calculation.
We focus on results for proton-proton collisions at 14 TeV centre-of-mass energy.
The inclusive event sample merging NLO QCD calculations
for 0, 1 and 2 additional jets is generated by using \madgraph\footnote{NLO inputs are generated with
aMC@NLO, with loose cuts and employing \textsc{Pythia} shower subtractions, and post-processed
by performing a single \textsc{Pythia} evolution step using the
settings recommended in the \href{http://amcatnlo.web.cern.ch/amcatnlo/list_detailed2.htm}{\madgraph documentation}.
The results after post-processing are stored and used as input for \textsc{Dire}.
LO events are also generated with loose cuts. Tighter merging scale cuts on LO and
NLO samples are applied by \textsc{Dire}. Table \ref{tab:nevents} lists
the number of events after the merging scale cut.}
merged and showered by \textsc{Dire}. Multiparton interactions, hadronisation
and beam remnants are handled by \textsc{Pythia}. The outputs are analysed using
\textsc{Rivet}~\cite{Bierlich:2019rhm}, which provides the input distributions to the Positive
Resampler, and performs the analysis and plotting. We present results for the
standard Rivet analyses \texttt{MC\textunderscore{}WINC} and \texttt{MC\textunderscore{}WJETS} in order to illustrate
the performance of the Positive Resampler on analyses already in use by the
community for testing the description of the production of a $W$-boson.

We then apply a series of resampling steps to the generated weighted
event samples. We first choose a target (binned) distribution which we aim to
preserve exactly during the procedure. We are free to pick an
arbitrary distribution for this purpose. In the present work, we consider
first the following one-dimensional
examples.
\begin{itemize}
\item[a)] \emph{W boson transverse momentum.} We preserve
  $\tfrac{d\sigma}{dp_{W\perp}}$, in bins with a width of 5\,GeV each. In
the peak region $p_{W\perp} < 125\,$GeV this coincides with a histogram from the
\texttt{MC\_WINC} Rivet analysis.
\item[b)] \emph{Parton shower evolution variable $t$.}
% \todo{Describe $t$ here or elsewhere.}
This unmeasurable parameter is defined as the \textsc{Dire} parton shower
ordering variable~\cite{Hoche:2015sya} at which a zero-parton state transitions
to a one-parton state. For events with more than zero partons before
showering, we use the $t$-value of the first node in the reconstructed
parton-shower history that is employed when calculating no-emission
probabilities within UNLOPS. If no transition from a zero- to a one-parton
state exists (e.g. if the shower did not perform an emission, or if the
history favoured electroweak clusterings), we set $t=-1$.
For $t>0$, we consider $\tfrac{d\sigma}{d\log t}$ in
$\sqrt{N}$ bins, where $N$ is the number of events in each sample. We
add one more bin for events with $t < 0$.
\item[c)] \emph{Total cross section.} All events are contained in a single bin.
\end{itemize}
Our aim here is to demonstrate the viability of positive resampling
independently of the concrete analysis. For this reason we generally
consider distributions for the complete event samples and refrain
from applying any analysis cuts at this stage. The exception is option
a), where the Rivet analysis is used for identifying the momentum of the W
boson.

In the next step, we apply the following procedure to each bin in the chosen
distribution.  These are the same steps later applied in each bin of
multi-dimensional sampling.
\begin{enumerate}
\item \emph{Change weights.} We turn negative-weight into
positive-weight events as described in section~\ref{sec:posres}. This
preserves the height of each bin exactly.
\item \emph{Partial unweighting.} We choose a target weight $W_t$ and
apply standard unweighting to all events with weights $w_t < W_t$,
i.e. we keep the event with probability $p = \tfrac{w_t}{W_t}$ and
then adjust the event weight to $W_t$. We choose $W_t$ in such a way
that approximately $10\%$ of the original events are kept. This number can
straightforwardly be adjusted.
\item \emph{Bin restoration.} We rescale the weights of all events in
the bin such that the original bin height is restored. If the
stochastic unweighting in the previous step resulted in an empty bin, we first
pick a random discarded event and add it back to the bin prior to
restoring the height. The probability to recover a discarded event is
chosen to be proportional to its weight.
\end{enumerate}
We also compare our approach to more traditional unweighting, where
the signs of the event weights are unchanged, and the modulus of the
event weight is used when deciding whether an event should be kept.
\begin{figure}[tbp]
  \centering
  \includegraphics[width=0.49\linewidth]{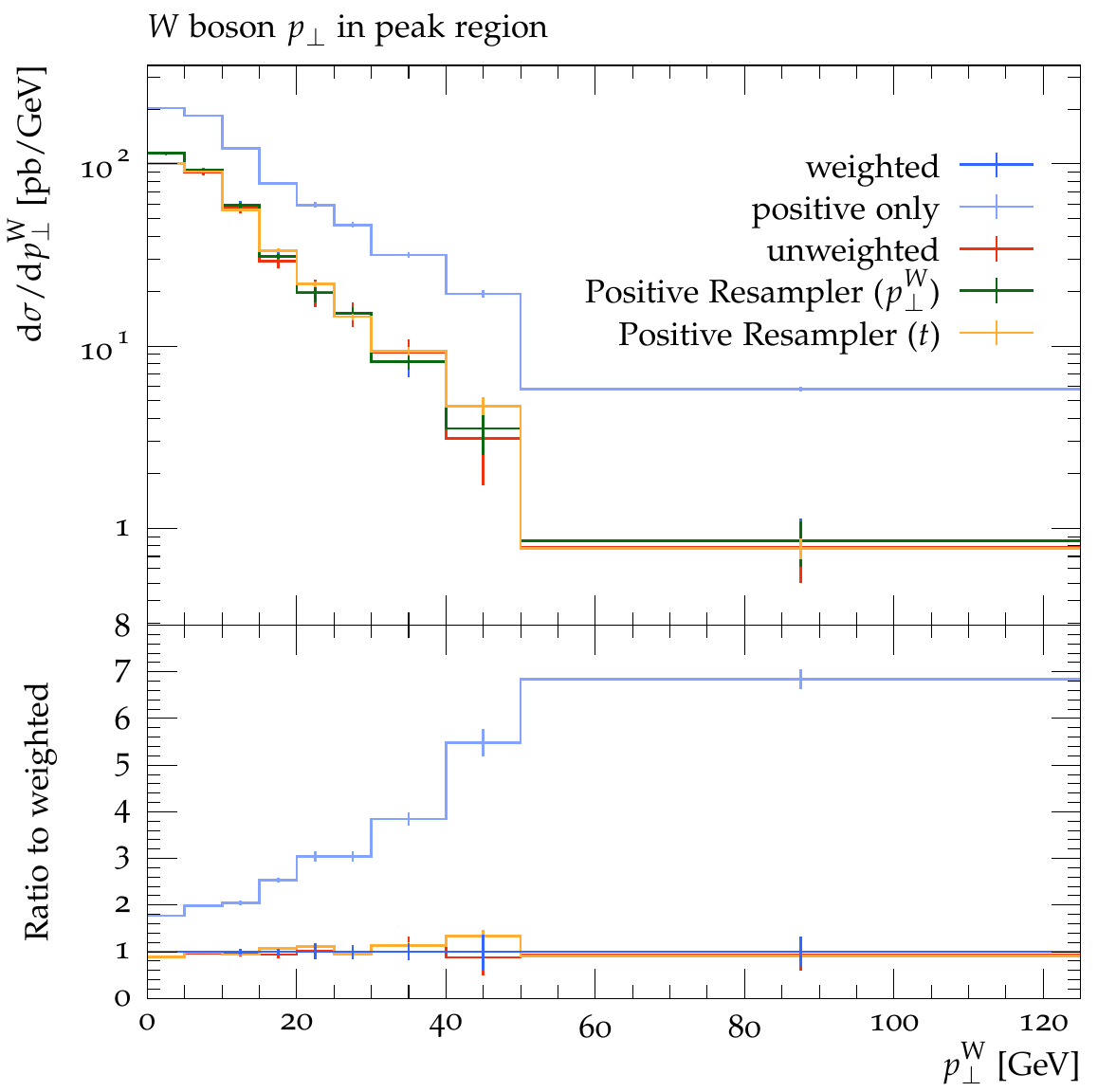}\hfill
  \includegraphics[width=0.49\linewidth]{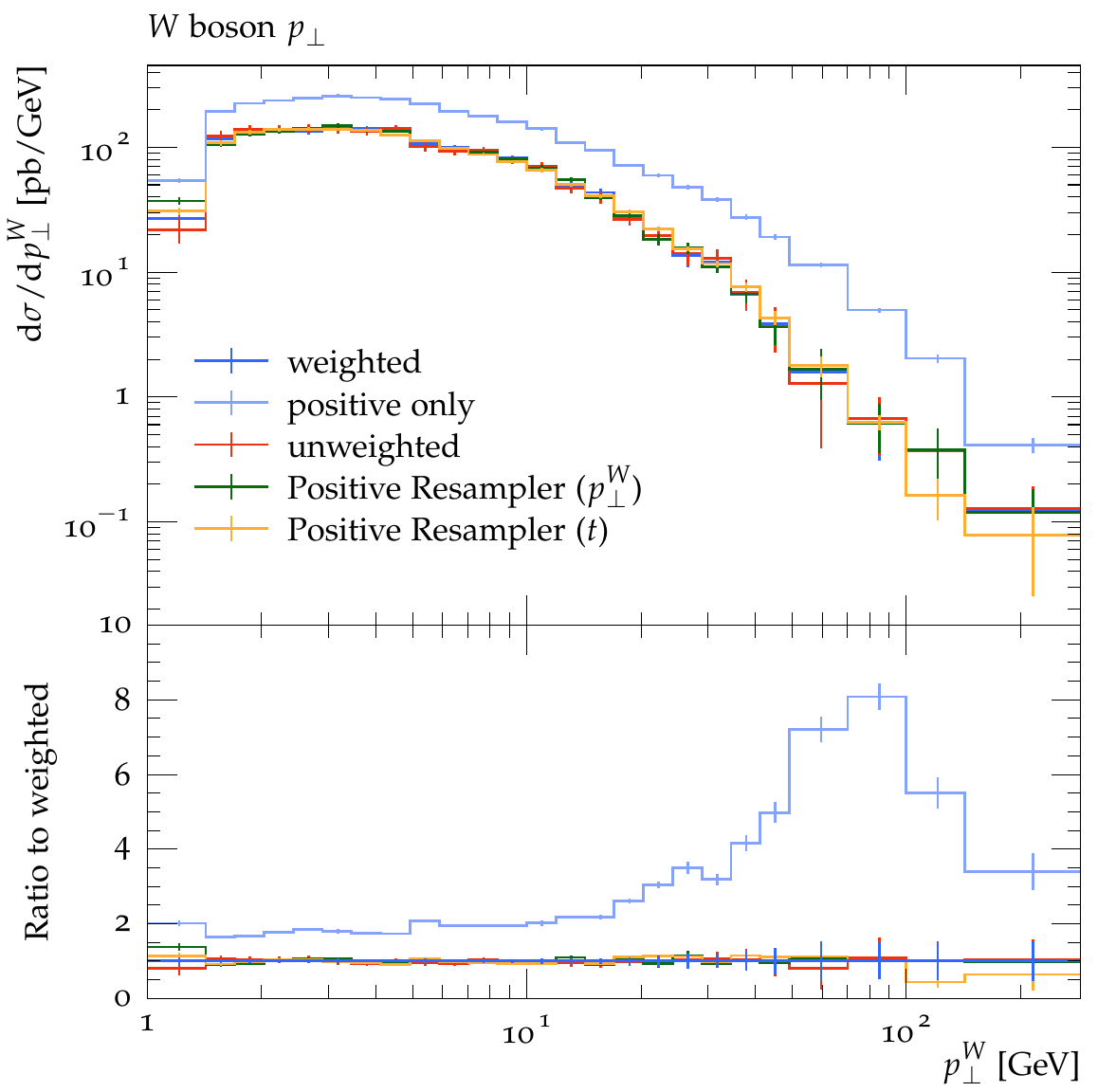}\\
  \includegraphics[width=0.49\linewidth]{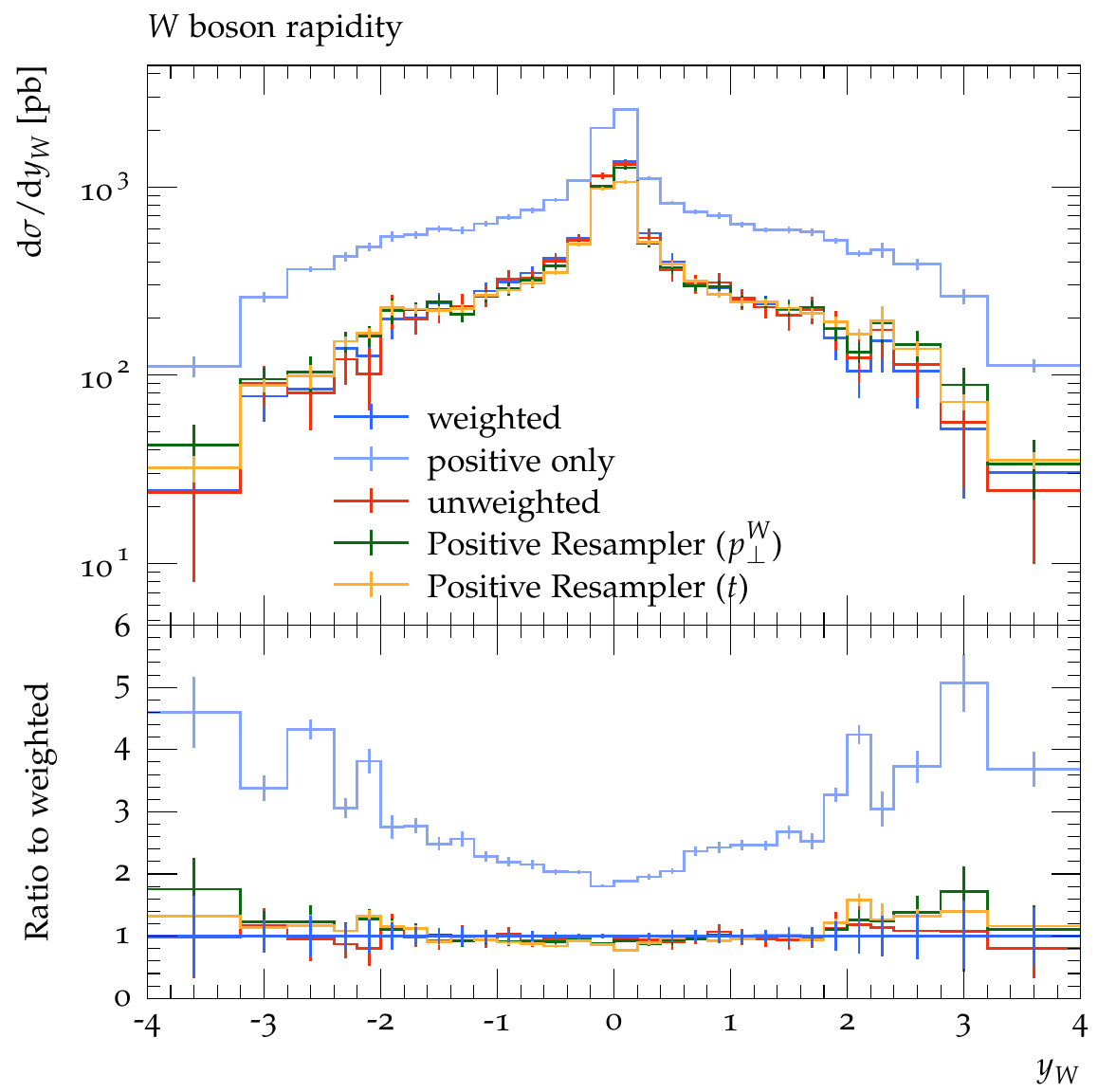}\hfill
  \includegraphics[width=0.49\linewidth]{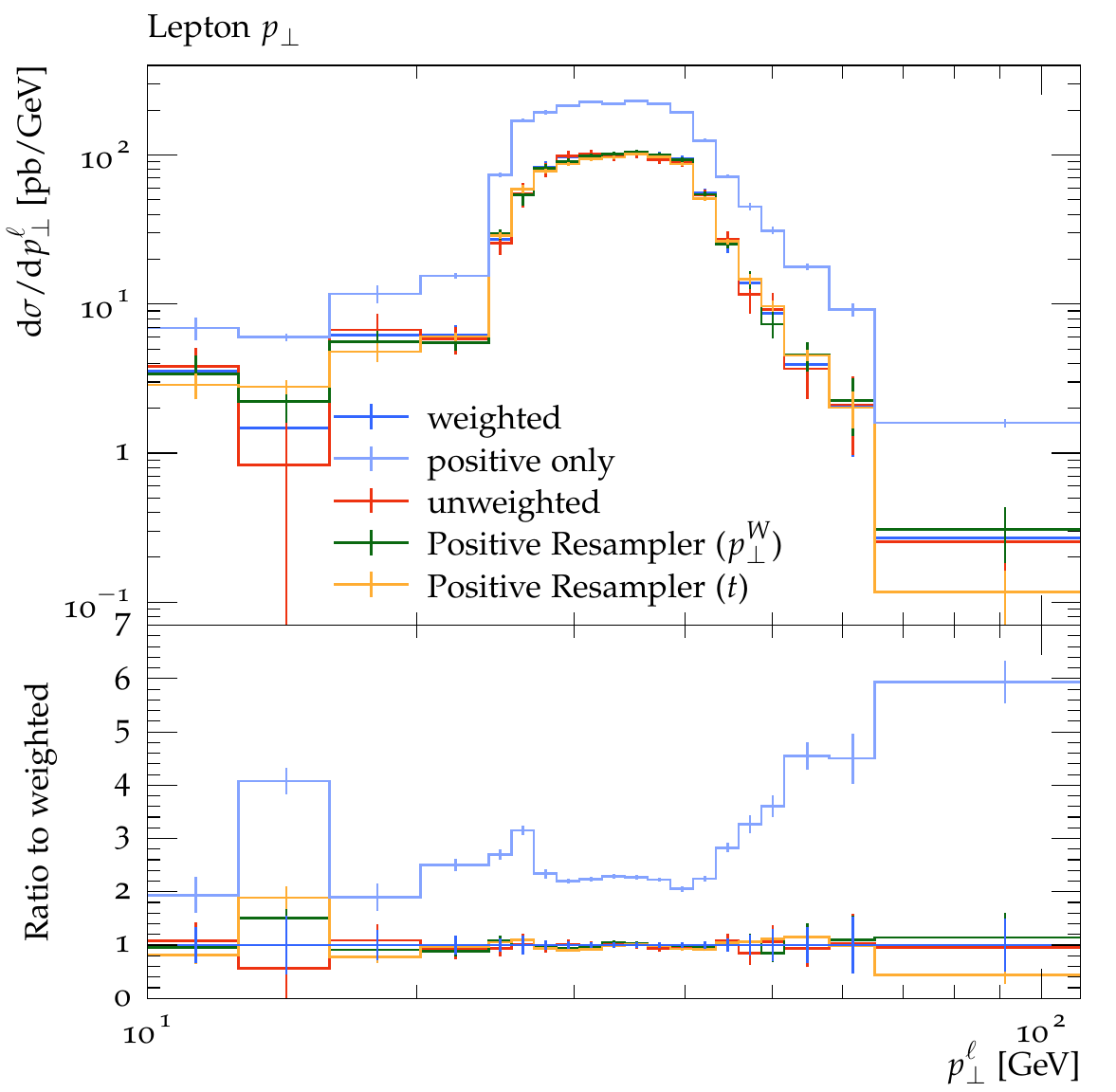}
  \caption{The distributions of $p_\perp^W$, $y_W$ and $p^l_\perp$ obtained
    using the various samples. The sample of ``positive only'' events is
    included to illustrate the scale of the contribution from negative weight
  events.}
  \label{fig:distributions}
\end{figure}
As is illustrated on Figure~\ref{fig:distributions}, the results
obtained using both traditional unweighting and the Positive Resampler are
consistent with the result using all events, well within the statistical
fluctuation of that largest sample. It is also clear that the statistical
uncertainty associated with the distributions obtained with the Positive
Resampler is similar to that of the input \emph{weighted} distribution. The
central values and associated uncertainty is for both calculated by Rivet.

To some degree, the power of the method is in the reduction of events
necessary to achieve a certain statistical accuracy.
\begin{table}[tbp]
  \centering
  \begin{tabular}{lrr}
    \toprule
    Sample&Total number of events&Events included in analysis\\
    \midrule
    weighted&5.3M&195k\\
    positive only & 3.2M&121k\\
    unweighted&1.5M&52k\\
    Positive Resampler($t$)& 659k&25k\\
    Positive Resampler($p_\perp^W$)& 33k&33k\\
    \bottomrule\\
  \end{tabular}
  \caption{The number of events generated in the sample of $W+0j,1j,2j$@NLO
    merged with UNLOPS (\emph{weighted}), and the number of events arising
    from these by using the algorithms discussed in the paper.}
  \label{tab:nevents}
\end{table}
Table~\ref{tab:nevents} lists the number of events in each sample; the
input test sample
to the Positive Resampler contains 5.3M events, of which
195k pass the cuts of the analysis. Of these, 3.2M and 121k respectively have
positive weights. The simple unweighting procedure leaves 1.5M (52k passing
cuts of the analysis). The Positive Resampler according to $t$ reduces this
to 659k and 25k respectively. The Positive Resampler($p_\perp^W$) by construction
accepts events only if they pass the cuts of the analysis, which with this
resampling is 33k. The true power of the positive resampling is in the
reduction from 52k to 25k or 33k of events. This reduction in the number of
events is brought about by the cancellation between positive and negative
contributions, illustrated in Figure~\ref{fig:distributions} by the
contribution of ``positive only''.  The fact that this contribution is
roughly twice the cross section at the peak of the distribution (most easily
checked for $y_w$) leads directly to the reduction in the number of events
obtained with the Positive Resampler. The increasing ratio of the ``positive
only'' to the ``weighted'' result at large $p^W_\perp$ and $p_\perp^l$
is a result of the increasing cancellations taking place within the input
sample as discussed in section~\ref{sec:unlops}. It is these cancellations that the Positive Resampler implements
effectively by reducing the event count. The number of events left
after the resampling can be adjusted and tuned -- the largest
possible number of events per unit of cross section is given by the number of
events (positive and negative) in the bin with the least cancellation,
divided by the cross section in this bin. The distributions expose large
cancellations in some regions of phase space, indicating that many more
events are required to obtain statistically meaningful spectra. Improvements
of the example NLO-merging implementation used -- to reduce the amount of
cancellation already at the ``weighted" stage -- would clearly be beneficial~\cite{Frederix:2020trv}.

\subsection{Positive Resampling in Higher Dimensions}
\label{sec:PosResHighDim}
\begin{figure}[tbp]
  \centering
  \includegraphics[width=0.49\linewidth]{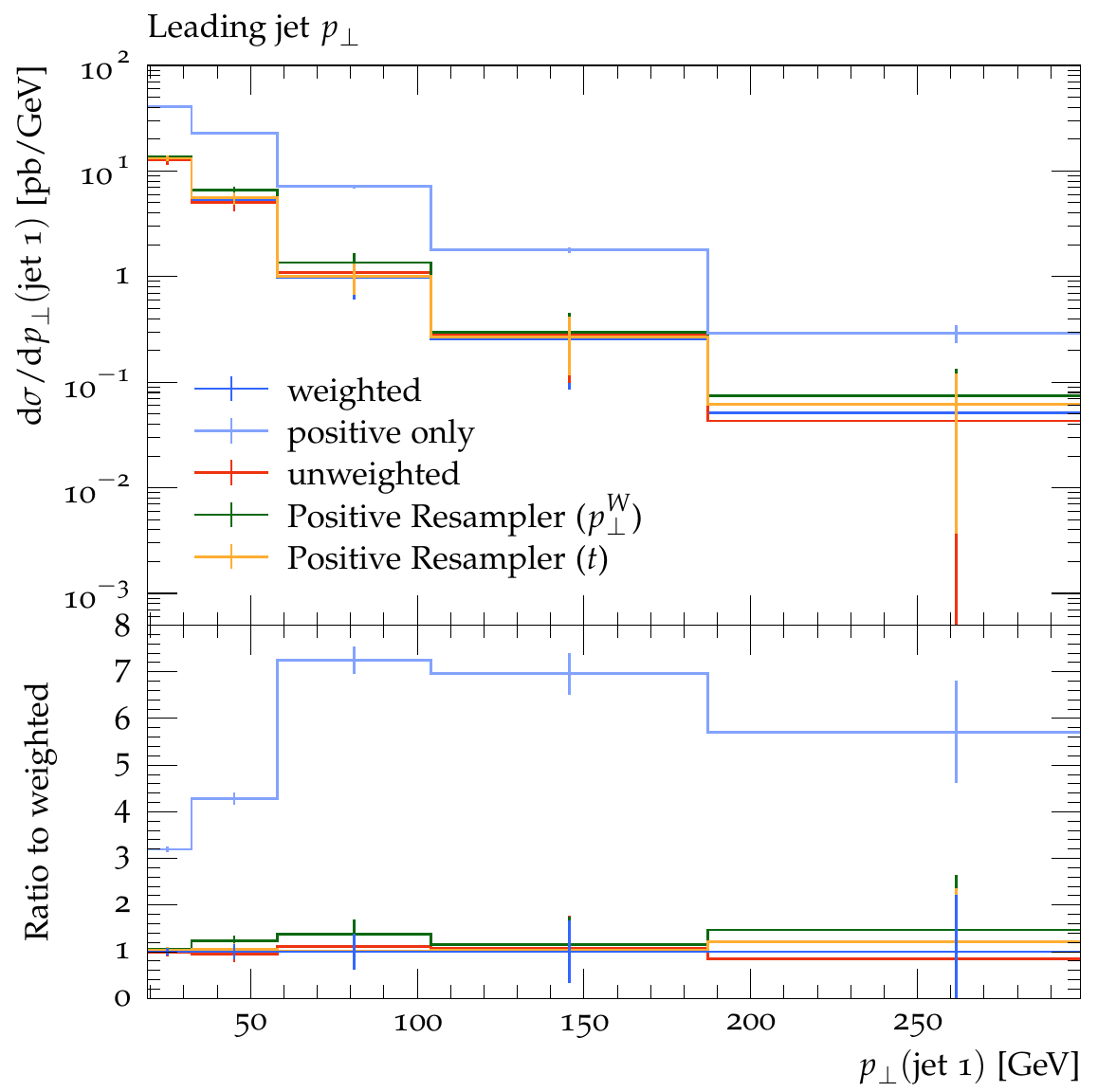}\hfill
  \includegraphics[width=0.49\linewidth]{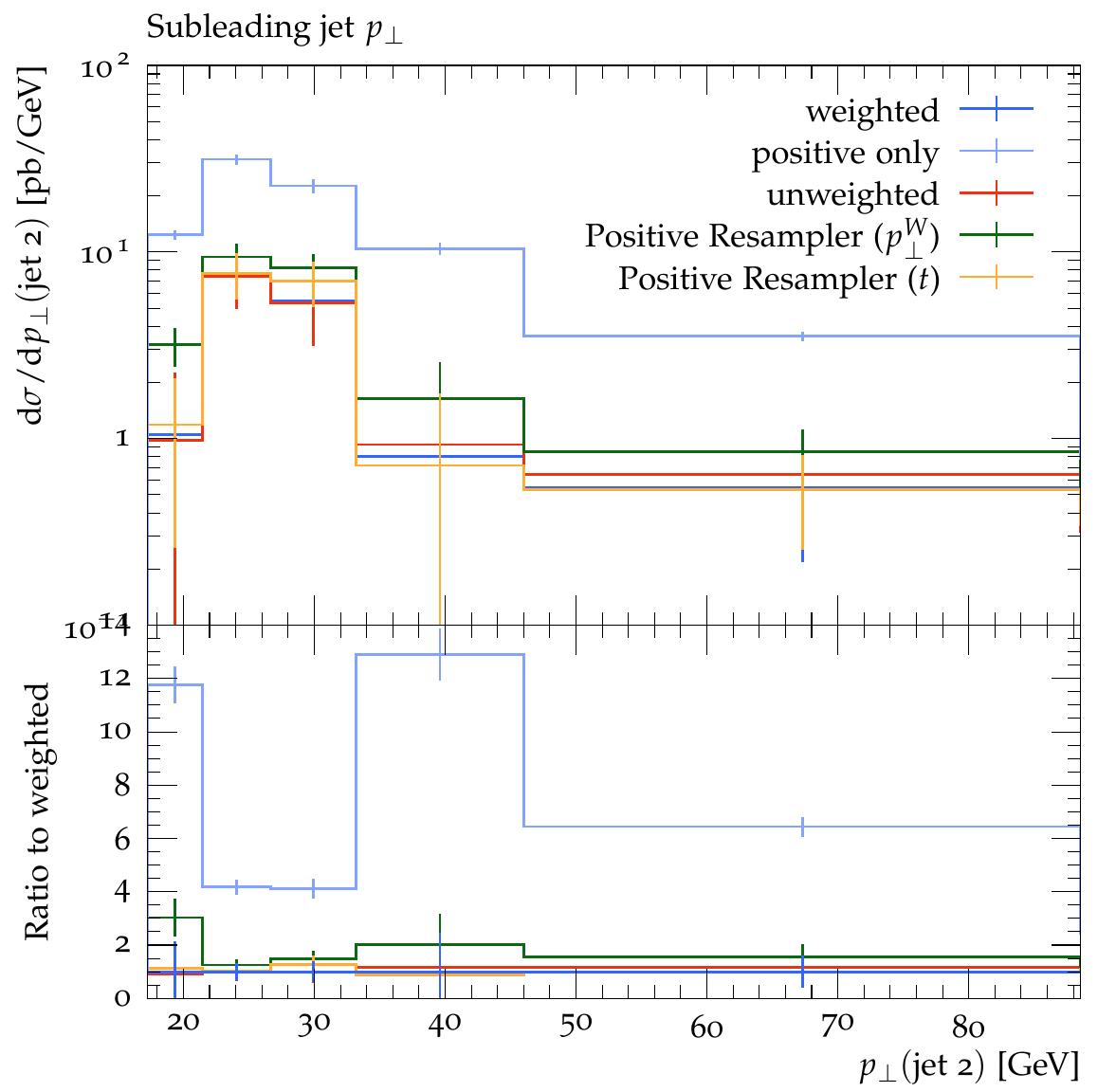}\\
  \includegraphics[width=0.49\linewidth]{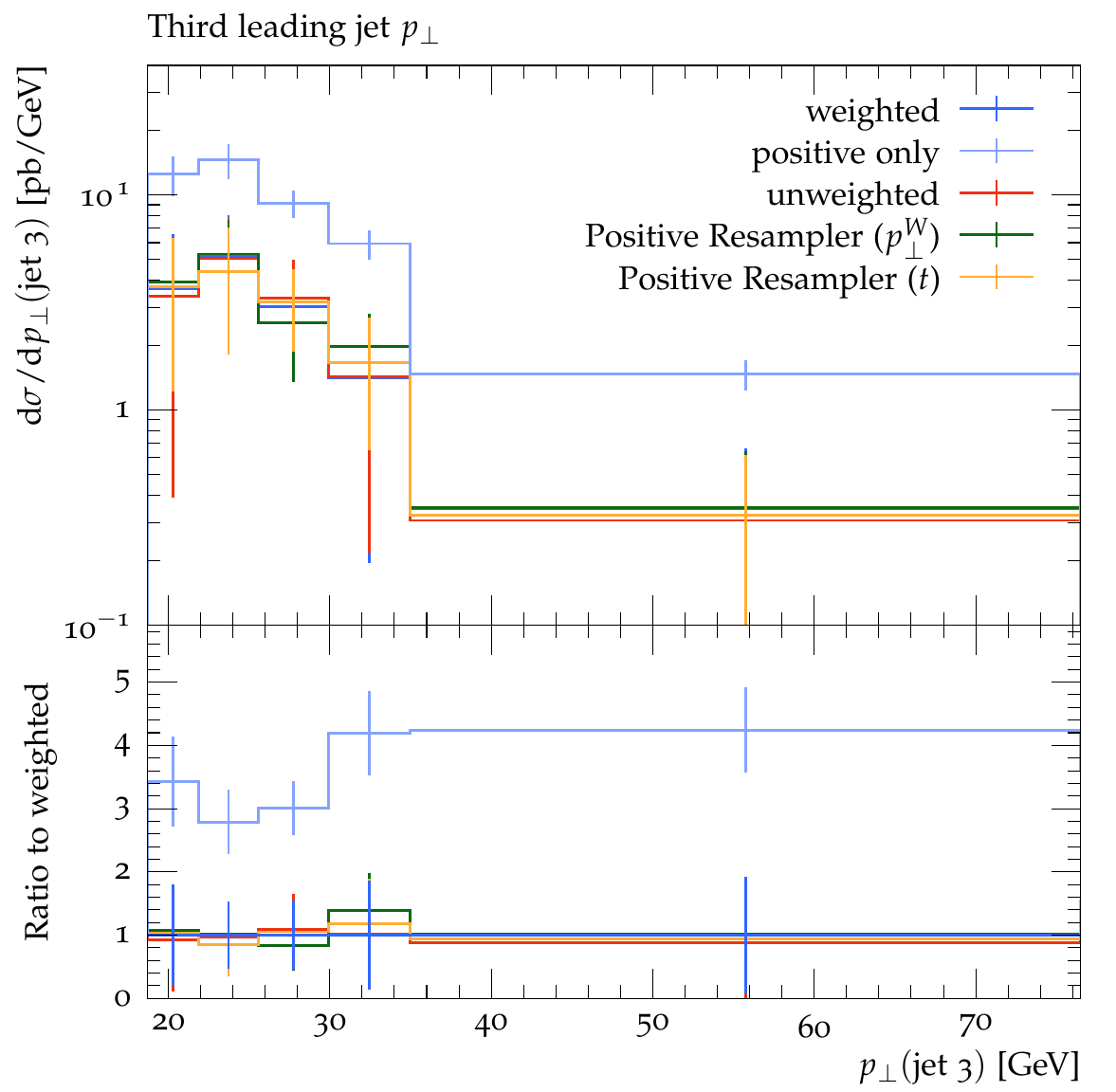}\hfill
  \includegraphics[width=0.49\linewidth]{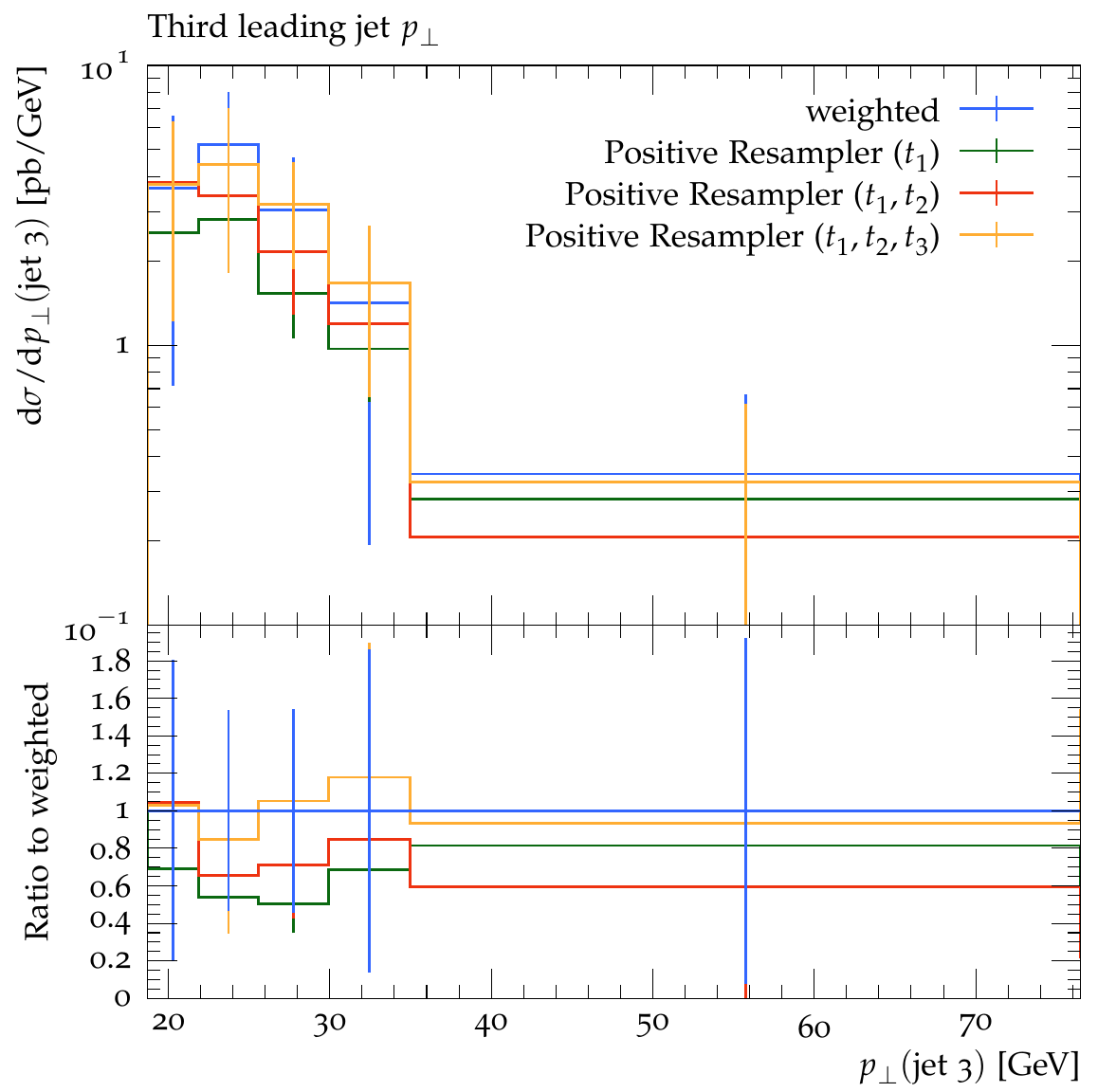}
  \caption{The transverse momentum of the hardest jet ($p^{j_1}_\perp$),
    second hardest jet ($p^{j_2}_\perp$), and third hardest jet
    ($p^{j_3}_\perp$). The lower right plot shows the convergence of the
    result of the Positive Resampler for an increasing number of dimensions
    using the shower evolution variable of up to three emissions.}
  \label{fig:pt123}
\end{figure}
The distributions studied in the previous section all relate to the momenta
of the $W$ and its decay products. Figure~\ref{fig:pt123} presents results
for the transverse momentum distribution of the three hardest jets produced
in association with the $W$. Following the discussion in
section~\ref{sec:multidimsampl} on multi-dimensional resampling for processes
with many final state momenta of leptons and jets , it is unsurprising that a
good description of the transverse momentum of the third jet requires
sampling in more than just one dimension. Perhaps the surprising result is
how \emph{few} extra dimensions are required for a good description. The
lower right pane on Figure~\ref{fig:pt123} shows the description of
$p_\perp^{j_3}$ when the Positive Resampler is applied in an increasing
number of dimensions from 1 to 3, using the shower evolution variable for the
first, second and third hardest emission. The result converges to the
\emph{weighted} result (the input to the resampler) as the number of
dimensions used in the resampling is increased, and using three dimensions it
is already well within the statistical fluctuations of the sample (some
statistical fluctuation is obviously expected from the stochastical process
of unweighting).

The result for the transverse momentum spectrum for the leading, subleading
and the third leading jet is presented on the three remaining plots of
Figure~\ref{fig:pt123}. The result for the Positive Resampler in $p_\perp^W$
uses just one-dimensional resampling, whereas that of \emph{Positive
  Resampler($t$)} uses the three-dimensional resampling in the shower
evolution variable. The curve for \emph{positive only} illustrates again the
significance of the negative weight events in the sample, and the portion of
the events that can be removed. Resampling in three scalar dimensions is
sufficient for the description of all the momenta.

\section{Conclusions}
We presented the \emph{Positive Resampler}, a method for modifying the
weights of events drawn from a sample with both positive and negative
weights, such that all events enter with positive weights, while
kinematic distributions and observables are preserved exactly or
within the statistical variations in the sample. The method was
demonstrated using reweighting in three different distributions and
the impact on 6 independent observables studied. Since weight cancellations are
handled explicitly, the reweighting can be implemented through
a reduction of the number of events, thus allowing to significantly lower
the cost and time associated with post-processing the event sample with
subsequent analysis and modelling. The implementation of
the reweighter used in this study will be publicly available after
publication of this manuscript.

\section*{Acknowledgements}
The ideas presented in this work arose from the inspiring environment and
discussions of the 2019 Les Houches Workshop ``Physics at TeV
Colliders''. The authors would like to express their gratitude to the
organisers for their continued efforts in creating the stimulating
milieu.

This work has received funding from the European Union's Horizon 2020
research and innovation programme as part of the Marie
Sk{\l}odowska-Curie Innovative Training Network MCnetITN3 (grant
agreement no. 722104), the EU TMR network SAGEX agreement No. 764850
(Marie Sk{\l}odowska-Curie), COST action CA16201: Unraveling new
physics at the LHC through the precision frontier, and by the Swedish
Research Council, contract number 2016-05996.

\bibliographystyle{JHEP}
\bibliography{papers}

\end{document}